\documentclass{article}  
\usepackage{breckenr2005}
\usepackage{hyperref}
\frompage{000} \topage{000}                                              
\def\rightmark{Tomographic Studies of the sQGP}
\def\leftmark{W. G. Holzmann for the PHENIX Collaboration}
 
\title{Tomographic Studies of the sQGP at RHIC} 
\authors{
{Wolf G. Holzmann$^1$ for the PHENIX Collaboration %
}\\[2.812mm]
{\normalsize
\hspace*{-8pt}$^1$ Department of Chemistry, State University of New York at Stony Brook, \\ 
Stony Brook, New York 11794-3400, USA\\[0.2ex] 
}}
 
\abstract{
Azimuthal correlation functions are used to study jet- and di-jet properties 
as a function of centrality in Au+Au collisions at 
$\sqrt{s_{_{\rm NN}}}$=200~GeV. Utilizing a novel technique to decompose the correlation function into 
a (di-)jet and an underlying event, the jet-pair distribution is extracted and 
compared to similar results for d+Au collisions obtained at the same collision energy.
A striking similarity is observed between the widths and associated yields of the (di-)jet 
distributions for d+Au and peripheral Au+Au collisions. By contrast, the distributions for 
mid-central Au+Au collisions indicate an increase in the di-jet yield with centrality, 
and a very broad away-side jet having a possible minimum at $\Delta\phi \approx \pi$. These 
features point to significant medium induced modification to the away-side jet and are compatible 
with recent predictions of jet-induced ``conical flow".
}
\PACS{specifications see, e.g.\ {\tt http://www.aip.org/pacs/}}
 
\begin{document}
 
\maketitle
\setcounter{page}{1}

\section{Introduction}\label{intro}

Reactions between Au ions at the Relativistic Heavy Ion Collider (RHIC), indicate the 
creation of a fireball of nuclear matter having energy density well above that required for 
a de-confined phase of quarks and gluons (QGP)~\cite{Adcox:2004mh,Fodor:2001pe}. The decay 
of this matter results in large azimuthal anisotropies in the particle emission patterns, suggesting 
early thermalization and the development of substantial pressure gradients which drive the
dynamical evolution of the system~\cite{Lacey:2001va,Shuryak:2004cy,Arkadij_ND,Gyulassy_qgp,Heinz,Heinz2}.
Hydrodynamic evolution of the fireball is further corroborated by the observation of 
strong radial flow~\cite{Masashi_NXu} and first hints of a long-range emitting source from a recent 
Imaging analysis~\cite{Lacey_ND,PChung_Praha}. Strong indications for hydrodynamic evolution of 
the emitting system implies the production of strongly interacting high energy density matter in 
energetic RHIC collisions~\cite{Shuryak:2004cy,Gyulassy_qgp,Heinz2}. Indeed, this matter has 
been observed to strongly suppress the yield of hadrons with large 
transverse momenta~\cite{ppg003,ppg014,star_supp,ppg023} and to suppress the away-side jet 
in central Au+Au collisions~\cite{Adler:2002ct}. It is believed that this suppression results
from energy loss of hard-scattered partons traversing the high energy density matter 
prior to the formation of hadrons~\cite{bjorken,appel1986,blaizot_mclerran86,wang2,gyulassy,wang}. 

	An important open question of great current interest is the influence of the 
parton-medium interactions on jet properties. Such an influence is of paramount importance 
if one wants to use jets as a probe of the properties of strongly interacting 
high energy density matter. Several recent works have outlined a possible influence of the 
coupling between jets and a strongly interacting 
medium~\cite{Ko,Hwa,Fries,Salgado,stoecker,stoecker2,shuryak,colorwake,Majumder:2004pt}. A particularly important 
proposal is the conjecture that the energy deposited in the medium could lead to the creation 
of a shock wave around the propagating parton, thereby creating ``conical flow'' or ``bow waves"
analogous to a sonic boom in a fluid\cite{stoecker,stoecker2,shuryak,colorwake}. The experimental observation of 
such conical flow  could serve to pin down the sound speed in the nuclear matter created 
at RHIC.

	In order to probe the influence of possible parton-medium interactions on jet properties, we 
use azimuthal angular correlation functions to investigate jet topologies and yields in d+Au and 
and Au+Au collisions. Here, the operational strategy is that d+Au measurements provide a good 
baseline for comparison to the Au+Au measurements which are expected to show much stronger modifications
to jet properties.

\section{Data Analysis}\label{techno} 
The analysis presented in this paper uses Au+Au and d+Au data ($\sqrt{s_{NN}}$=200 GeV) 
provided by RHIC in the second and third running periods (2001,2003), respectively. The full PHENIX detector 
setup is described elsewhere~\cite{nim_1}. Charged tracks relevant to this analysis 
were reconstructed in the central arms of PHENIX, each of which 
covers 90 degrees in azimuth. Tracking was performed via the drift chamber and 
two layers of multi-wire proportional chambers with pad readout (PC1,PC3)\cite{nim_1}. 
A combinatorial Hough transform in the track bend plane was used for pattern 
recognition\cite{nim_2}. Most conversions, albedo and decays were
rejected by requiring a confirmation hit within a 2 $\sigma$ matching window in the PC3.
Collision centrality was determined via cuts in the space of BBC versus ZDC analog 
response \cite{nim_3}.

The correlation function in relative azimuthal angle between particle pairs, $\Delta \phi=(\phi_{1}-\phi_{2})$,
is defined as the ratio of two distributions
\begin{equation}
C\left(\Delta\phi\right) \propto \frac{N_{cor}\left( \Delta\phi
\right)} {N_{mix}\left( \Delta\phi \right)}.
\end{equation}
 The foreground distribution $N_{cor}$, measures coincident particle pairs from 
the same event by pairing particles from a high-$p_{T}$ ``trigger" bin 
($2.5 < p_T < 4.0$~GeV/c, hereafter labeled A) 
with associated particles from a lower $p_{T}$ selection ($1.0 < p_T < 2.5$~GeV/c, hereafter labeled B). 
The background distribution $N_{mix}$, is generated in an analogous way by mixing particle pairs 
from different events within the same multiplicity and vertex class. The 
Azimuthal acceptance and detector efficiency effects cancel in the ratio of foreground to
background distributions and the correlation function yields the probability distribution for detecting 
correlated particle pairs per event within the PHENIX 
pseudorapidity acceptance ($|\eta|<0.35$).
\begin{figure}[hbt!]
\begin{center}
\vspace{-.1in}
\epsfysize 0.8\textwidth
\epsffile{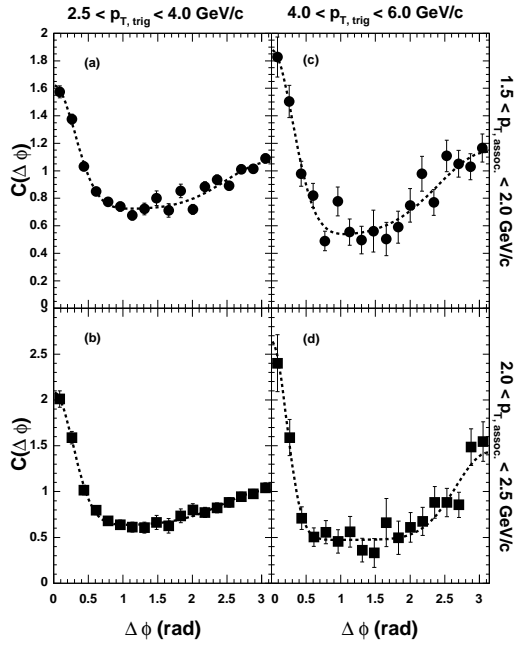}
\caption[]{d+Au correlation functions $C(\Delta \phi)$ for several $p_{T}$ selections
as indicated. The centrality selection is 0-80\%.
The dashed line represents a double gaussian fit to the data.\label{Fi:fig0}}
\end{center}
\end{figure}
Fig.~\ref{Fi:fig0} shows d+Au correlation functions for two different trigger and associated 
$p_T$ selections as indicated. The centrality selection is 0-80\%.
The dashed line represents a double Gaussian fit to the data.
These d+Au correlation functions exhibit a shape reminiscent of what one would expect 
from di-jet fragmentation. That is, a relatively narrow near-side peak centered at $\Delta\phi=0$ 
and a somewhat wider away-side peak centered at $\Delta\phi=\pi$. Fig.~\ref{Fi:fig0} shows that 
both near-side and away-side peaks narrow and are more pronounced for higher $p_T$-selections 
of trigger and associated particles, respectively. Such a pattern is expected if (di-)jet
fragmentation is the dominant particle production mechanism at high transverse momenta. 
It is noteworthy that the widths (of near- and away-side jets) and the yields obtained 
from d+Au correlation functions are rather similar to those obtained from p+p collisions. 
Consequently, we ascribe all correlation in d+Au collisions to jets. 

\begin{figure}[hbt!]
\begin{center}
\vspace{-.1in}
\epsfysize 0.85\textwidth
\epsffile{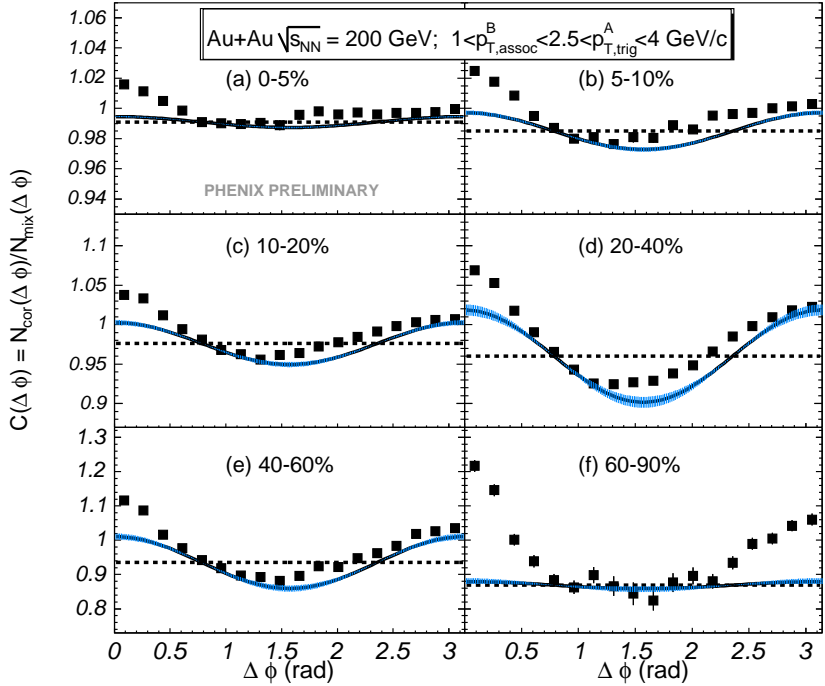}
\caption[]{Correlation functions $C(\Delta\phi)$, generated for trigger particles in 
2.5~GeV/c$<p_{T}^{A}<$4.0~GeV/c and associated particles in 1.0~GeV/c$<p_{T}^{B}<$2.5~GeV/c. The bands show 
the harmonic contribution within the systematic uncertainty. 
The dotted lines indicate the value of $a_o$ (see text).\label{Fi:fig1}}
\end{center}
\end{figure}
The d+Au correlation functions (cf. Fig.~\ref{Fi:fig0}) are to be compared to the correlation
functions for Au+Au data shown in Fig.~\ref{Fi:fig1}.
The filled squares in the figure show correlation functions for AB charged hadron 
pairs for several indicated centralities. It is difficult to overlook the striking similarity 
between the correlation function obtained for the most peripheral collisions 
(cf. Fig.~\ref{Fi:fig1}f) and that obtained for d+Au collisions (cf. Fig.~\ref{Fi:fig0}). Both 
correlation functions show the two narrow peaks (located at $\Delta\phi=0$ and $\Delta\phi=\pi$)
characteristic of (di-)jet fragmentation. By contrast, the correlation functions for 
more central Au+Au collisions show strong indications for a harmonic component and 
hints for a broad away-side jet ie. 
the correlation functions show a minimum below $\Delta \phi = \pi/2$. Such a shift away from the 
minimum expected for harmonic contributions, can only come about if the away-side jet is 
significantly broader than that observed in d+Au collisions. An important finding that should 
be stressed here is that the observed characteristics of all of the Au+Au correlation functions 
can be fully accounted for via two contributions to the correlation function: (i) a (di-)jet and 
(ii) a harmonic contribution~\cite{Adler:2002ct,Ajitanand:2002qd,Chiu:2002ma,Adler:2002tq}.

\section{Decomposition of jet and harmonic contributions}
	Careful investigation of possible modifications to the di-jet distributions in 
Au+Au collisions require access to the jet-pair distribution without the blurring effects 
of the harmonic contributions. Consequently, a reliable procedure for decomposing the 
measured correlation functions into their (di-)jet and harmonic (or flow) contributions are 
required. Detailed descriptions of such a procedure are given in 
Refs.~\cite{Stankus_RP,Ajit_Methods}. We give here only an outline of the main points.

It can be shown~\cite{Stankus_RP} that the pair correlations from the combination of 
flow and jet sources is given by 
\begin{equation}
C^{AB} (\Delta \phi ) = a_{o} [C^{AB}_H(\Delta \phi )] + J (\Delta \phi ), 
\label{eq1}
\end{equation}
where $C^{AB}_H(\Delta \phi )$ is a harmonic function of effective amplitude 
v$_{2}$, 
\begin{equation}
C_{H}^{AB} (\Delta \phi )_{ } = [1 + 2 v_{2} 
cos 2(\Delta \phi _{ }) ];  \  v_{2}=(v_2^A \times v_2^B).
\label{eq2}
\end{equation}
and J($\Delta \phi )$ is the (di-)jet function. 
No explicit or implicit assumption is required for the functional form of 
J($\Delta \phi$). Rearrangement of Eq.~\ref{eq1} gives
\begin{equation}
J(\Delta \phi ) = C^{AB}(\Delta \phi ) - a_{o}C^{AB}_{H} (\Delta \phi ). 
\label{eq3}
\end{equation}
Thus, one only requires knowledge about a$_{o}$ and v$_{2}$ to evaluate J($\Delta \phi$). To constrain, a$_{o}$
we assume that the (di-)jet function has zero yield at the minimum (ZYAM) $\Delta \phi _{min}$, 
in the jet function, i.e. $a_0C_{H}^{AB}(\Delta{\phi}_{min}) = C^{AB}(\Delta \phi_{min})$. This fixes the value 
of $a_{o}$.

	The $v_2$ value reflects the average anisotropy of the particles from both sources, and 
can be obtained from the single particle distributions relative to  
the reaction plane $\psi_R$. However, this step requires that 
the reaction plane is itself determined by a procedure essentially free of non-flow effects.
This is accomplished in the present analysis by demanding a large (pseudo)rapidity 
gap ($\Delta \eta \sim 3-3.9$) between the reaction plane and the particles correlated 
with it~\cite{Adler:2003kt,Adler:2004cj}. 
It is expected that the $v_2$ values so obtained are much less affected by jet  
contributions~\cite{Adler:2003kt}. In addition, the reaction plane (at each centrality)
and its dispersion correction, and $v_2^A$ and $v_2^B$ were 
obtained from the same data set used for jet function extraction in order to 
avoid potential biases. Correction for reaction plane dispersion followed the 
procedures outlined in Ref.~\cite{Poskanzer:1998yz}.
\subsection{Simulation tests}
Prior to applying the decomposition method to PHENIX data, its reliability was thoroughly tested 
via extensive Monte Carlo simulations~\cite{Ajit_Methods}. These investigations included 
simulation tests which took account of the $\phi$ and the $\eta$ acceptance of PHENIX. 
\begin{figure}[hbt!]
\begin{center}
\vspace{-.1in}
\epsfysize 0.85\textwidth
\epsffile{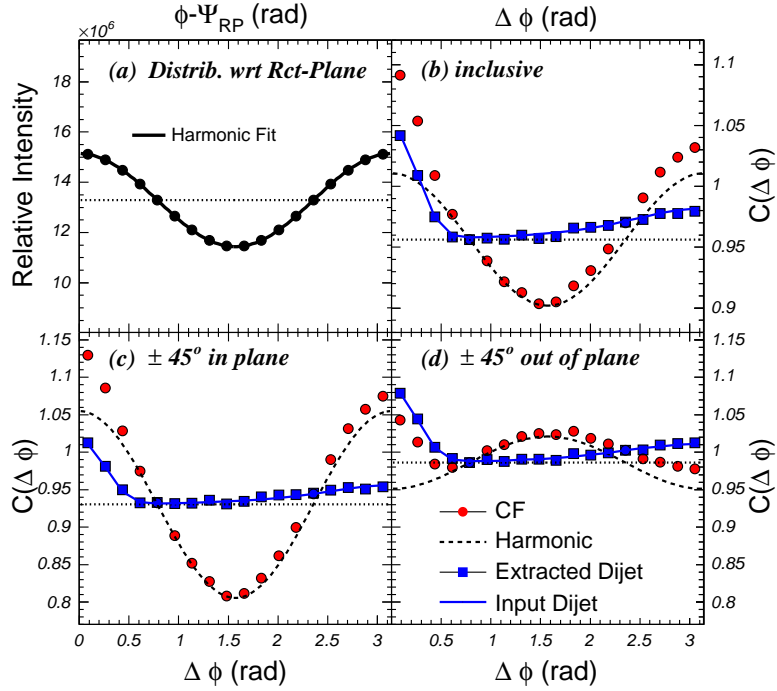}
\caption[]{Simulation test of the ZYAM decomposition procedure from \cite{Ajit_Methods}.
(a) simulated azimuthal distribution of particles with respect to the event plane. (b)
Simulated correlation function with harmonic component and extracted jet-pair distribution.
The solid line represents the input jet-pair distribution (referenced to $a_{o}$)
as obtained from tagging the jet particles in the simulation. (c) and (d) same as (b) but for
trigger particles constrained in-plane and out-of-plane respectively (see text).
\label{Fi:figsim}}
\end{center}
\end{figure}
Representative results from these Monte Carlo investigations are summarized in 
Fig.~\ref{Fi:figsim}. Panel (a) depicts the azimuthal distribution of the simulated 
reaction products with respect to the event plane. The smooth line is a harmonic fit to 
the data to extract $v_2$. This $v_2$ determines the amplitude of the harmonic component 
(dashed line in panel (b)) to be subtracted from the correlation function 
(filled circles) shown in panel (b). The solid squares in panel (b) shows the 
ZYAM subtracted jet-pair distribution referenced to $a_{o}$. This distribution is to be compared 
to the input jet pair distribution (solid line) obtained via tagging of the jet-particles 
in the simulation. The rather good agreement shown between input and output jet-pair distributions 
in Fig.~\ref{Fi:figsim}b serves to confirm the reliability of the method. The method is easily 
generalized to the case in which trigger particle detection is constrained within a cut angle 
parallel or perpendicular to the reaction plane~\cite{Ajit_Methods}. Results from simulations in 
which such constraints have been applied are summarized in panels (c) and (d) of Fig.\ref{Fi:figsim}. 
In this case, the harmonic function is determined following the techniques outlined 
in \cite{Voloshin}. Here again, panels (c) and (d) clearly indicate that the input jet function 
is reproduced in detail. It is noteworthy that a wide range of tests for a variety of input 
jet-pair distributions including those that might be expected from conical flow, were made with 
equally good recovery of the input jet-pair distributions~\cite{Ajit_Methods}. 
\subsection{Decomposition of the measured correlation functions}
The solid bands in each panel of Fig.~\ref{Fi:fig1} illustrate the 
application of the ZYAM condition to PHENIX data with the measured values of 
$v_2$ ($v_{2}=(v_2^A \times v_2^B$)).
The dashed lines show the $a_o$ value obtained for each centrality. Following Eq.~\ref{eq3}, the 
jet-pair distribution is obtained at each centrality via subtraction of the harmonic 
contribution from the correlation function. It is straightforward to show that 
the integral of this distribution is related to the average fraction of jet-correlated 
particle pairs per event and hence the conditional per trigger yield~\cite{Stankus_RP,Ajit_Methods}.
The ratio of the sum of $J(\Delta \phi)$ and the sum of $C(\Delta \phi)$ 
(over all bins in $\Delta \phi$) gives the fraction of jet-correlated particle 
pairs per event $PF$,
\begin{equation}
PF = \frac{\sum_i{J(\Delta\phi_i)}}{\sum_i{C(\Delta\phi_i)}}
\label{Eq:JPF_decomp}
\end{equation}
Subsequent multiplication of this fraction by the average number of detected particle 
pairs per event $\langle N^{AB}_{d} \rangle$, followed by a division by the product 
of the detected singles rates $\langle N^{A}_{d}\rangle$, $\langle N^{B}_{d}\rangle $, 
gives the event averaged jet-pair production in excess of the combinatoric pair production. 
A final product with the efficiency corrected singles rate $\langle N^{B}_{eff}\rangle$, for 
bin B, gives the efficiency corrected pairs per trigger or conditional 
yield $CY$~\cite{Stankus_RP,Ajit_Methods},
\begin{equation}
CY = PF \times \frac{\langle N^{AB}_{d} \rangle}{ \langle N^{A}_{d}\rangle \times \langle N^{B}_{d}\rangle}
\times \langle N^{B}_{eff}\rangle .
\label{Eq:CY_decomp}
\end{equation}

\section{Results}\label{others}
%
%
%
%
%
\begin{figure}[hbt!]
\begin{center}
\vspace{-.1in}
\epsfysize 0.85\textwidth
\epsffile{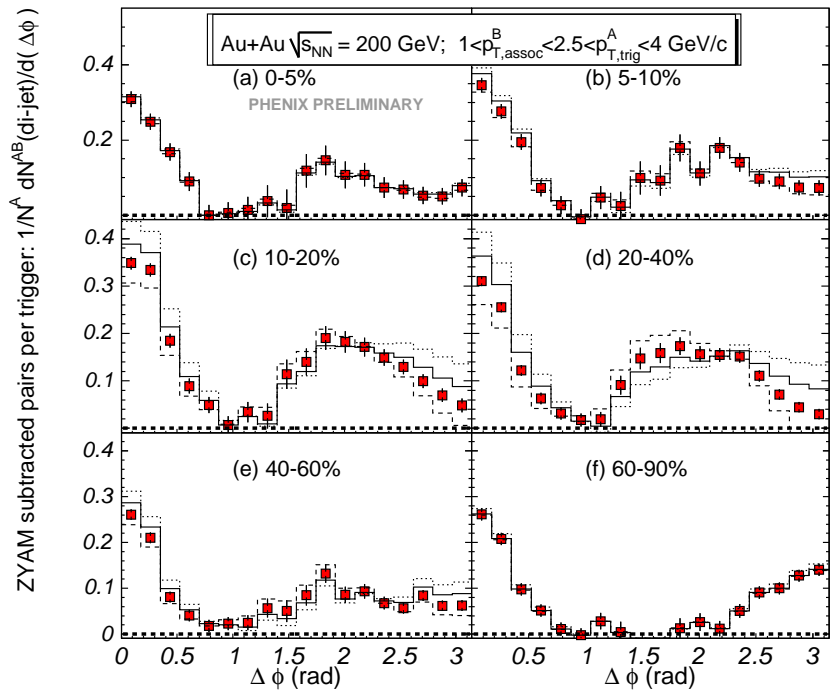}
\caption[]{ZYAM-subtracted jet-pair distributions $1/N_{trig}dN/d(\Delta\phi)$. The dashed(solid) histograms indicate the distributions resulting from increasing(decreasing) $v_2 = v_{2}^{A} \times v_{2}^{B}$ by one interval of the systematic error.\label{Fi:fig2}}
\end{center}
\end{figure}
The ZYAM-subtracted conditional yield distributions, normalized to the number of jet-pairs 
per trigger particle, are shown in Fig.~\ref{Fi:fig2}.
The distribution for the 60-90\% peripheral event sample (cf. Fig.~\ref{Fi:fig2}f) 
shows the typical (di-)jet shape that is familiar from p+p and d+Au collisions at RHIC.  
It consists of two distinct peaks, one narrow near-side peak centered at $\Delta\phi=0$ 
and a broader away-side peak at $\Delta\phi=\pi$. It is interesting to trace the 
evolution of these peaks with collision centrality, as one expects to make increasing 
amounts of hot and dense matter in the more central collisions.
Scanning the mid-central and central jet-pair distributions (Fig.~\ref{Fi:fig2}a-e), one 
observes that the near-side peak topology remains essentially unmodified. We attribute this 
to a possible trigger bias of the near-side jet. However, it should be noted that the 
near-side yield does show a mild rise with centrality as discussed below. In contrast to 
the near-side peak, the away-side peaks show indication for strong modifications (both in 
magnitude and in shape) for all centralities other than the most peripheral event 
selection. More precisely, in the 0-5\%, 5-10\% and 40-60\% samples, the away-side peak 
evidences a plateau like shape which is decidedly non-gaussian and much broader in 
width than that for the 60-90\% sample. For the 10-20\% and 20-40\% samples 
(Fig.~\ref{Fi:fig2}c-d), the away-side peak remains broad 
but also indicates an apparent local minimum at $\Delta\phi=\pi$ and a 
maximum at $\Delta\phi=2\pi/3$. This latter pattern is similar to recent
predictions of jet-induced "conical flow" \cite{stoecker,stoecker2,shuryak}. 
It should be pointed out however that these results do not preclude an alternative 
scenario which conjectures the combined influence of energy loss and the inclination 
angle of the jet with the flow field~\cite{Salgado}. Nonetheless, both approaches require 
relatively strong coupling between jets and the high energy density medium.

	The solid (dashed) lines in Fig.\ref{Fi:fig2} represent the conditional yield distributions
that would result from subtracting out a $v_2$ product lowered (raised) by one systematic 
error, respectively. The systematic error on $v_2$ is dominated by the uncertainty on the 
reaction plane dispersion. The dotted line indicates the jet-pair distribution resulting from a 
subtraction with $v_2$ product lowered by twice the systematic uncertainty. From these curves, 
one can see that a decrease in $v_2$ by two intervals of the systematic error can 
recover the local minimum at $\Delta\phi=\pi$. However, the result of a broadened away-side jet 
remains robust and the away-side jet-shapes remain non-Gaussian. Several studies are currently 
underway to firm up the mechanism/s responsible for the atypical away-side jet topologies observed 
in Au+Au collisions.
\begin{figure}[hbt!]
\begin{center}
\vspace{-.1in}
\epsfysize 0.55\textwidth
\epsffile{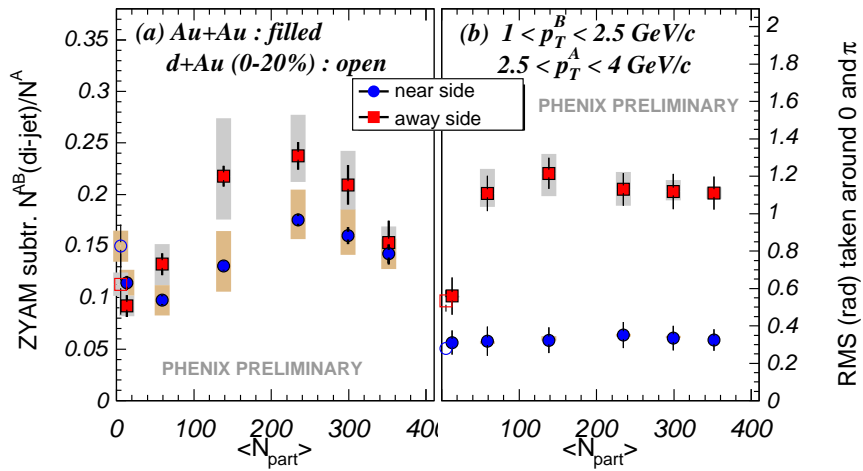}
\caption[]{(a)~Associated yields for near- and 
away-side peaks in the jet-induced pair distribution;
and (b)~Widths (RMS) of the peaks. The open symbols
denote results for 0-20\% most central d+Au events from
a recent PHENIX analysis.\label{Fi:fig3}}
\end{center}
\end{figure}

	To better quantify these jet properties, we split the jet-pair distributions into a near-side 
range, 0~-~$\Delta \phi_{min}$ and an away-side range  $\Delta \phi_{min}$ - $\pi$. 
The near- and away-side parts of the distribution are then further characterized by 
their RMS (taken around 0 and $\pi$) and their yield of associated pairs per trigger. 
These results are summarized in Fig.\ref{Fi:fig3} as a function of centrality. For comparison,
similar results are included for the 0-20\% most central d+Au collisions (open circles) obtained at 
$\sqrt{s_{NN}}=200 GeV$. The systematic uncertainty in $v_2$ has been propagated into the 
systematic errors for yields and RMS values. The systematic error on the yields also accounts for 
the systematic uncertainty on the single particle reconstruction efficiency. 

	Fig.\ref{Fi:fig3}b shows that both the near- and away-side widths for peripheral (60-90\%) Au+Au 
collisions compare well with those obtained for d+Au collisions. This is consistent with the expectation of 
very little, if any, medium induced modifications to jet-topologies in peripheral Au+Au 
collisions. An inspection of the centrality dependence of the near-side RMS shows no apparent 
change with centrality. On the other hand, the away-side width is significantly broadened 
for all but the most peripheral event sample, possibly indicating strong modifications to the fragmentation 
process by the hot nuclear medium. Although the near-side widths are centrality independent, 
Fig.\ref{Fi:fig3}a points to a mild increase in the near-side conditional yield from peripheral to 
central Au+Au collisions, possibly indicating that even the near-side fragmentation process might 
still be influenced by the medium. The apparent differences in the evolution of near- and away-side 
jet characteristics could be signaling  the contribution of several different mechanisms to 
jet-modification. 
%
\section{Conclusions}\label{concl}
In summary, we have used a novel correlation function technique to decompose jet correlations 
from collective long range harmonic correlations (elliptic flow). We find, that the extracted 
jet-pair distributions show a strong centrality dependent change in shape and associated 
per-trigger yield, especially for the away-side jet. The jet-pair distributions obtained for peripheral 
Au+Au collisions are very similar to those obtained for d+Au collisions but the distributions for more 
central Au+Au collisions are markedly different and are qualitatively consistent with several recent 
theoretical predictions of possible modification to jet fragmentation by a strongly interacting 
medium \cite{shuryak,stoecker,stoecker2,colorwake,Salgado}. Further experimental and theoretical studies are 
clearly required to establish the detailed mechanism/s responsible for the observed jet modification/s and 
to pin down the properties of the high energy density strongly interacting matter produced in energetic 
Au+Au collisions at RHIC. Several such studies are currently being pursued with vigor.

%
\section*{Acknowledgments}
I would like to thank the organizers for giving me the opportunity to present these results
in such a stimulating and friendly environment. I am deeply grateful to the Nuclear Chemistry Gang
at Stony Brook for much more than I have space left to write. Finally, these acknowledgments would not be complete
without a big Thank You to the PHENIX collaboration and the RHIC team.
 

\vfill\eject
\end{document}